\newif{\ifjournal}
  \journalfalse

\ifjournal
  \documentclass{aa}
  \usepackage{graphicx,times}
\else
  \documentclass{paper}
  
  \newcommand{\ga}{\gtrsim}
\fi

\renewcommand{\d}{\mathrm{d}}

\begin{document}

\title
  {Cluster detection from surface-brightness fluctuations in SDSS
   data}
\ifjournal
  \author
    {Matthias Bartelmann and Simon D.M.~White}
  \institute
    {Max-Planck-Institut f\"ur Astrophysik, P.O.~Box 1317, D--85741
     Garching, Germany}
  \authorrunning{M.~Bartelmann \& S.D.M.~White}
  \titlerunning{Cluster detection in SDSS data}
\else
  \author
    {Matthias Bartelmann and Simon D.M.~White\\
     Max-Planck-Institut f\"ur Astrophysik, P.O.~Box 1317, D--85741
     Garching, Germany}
\fi
\date{\today}

\newcommand{\abstext}
  {Galaxy clusters can be detected as surface brightness enhancements
   in smoothed optical surveys. This method does not require
   individual galaxies to be identifiable, and enables clusters to be
   detected out to surprisingly high redshifts, as recently
   demonstrated by the Las Campanas Distant Cluster Survey
   (LCDCS). Here, we investigate redshift limits for cluster detection
   in the Sloan Digital Sky Survey (SDSS). Calibrating assumptions
   about the surface brightness profile, the mass-to-light ratio, and
   the spectral energy distribution of galaxy clusters using available
   observational data, we show that it should be possible to detect
   galaxy groups out to redshifts of $\sim0.5$, and massive galaxy
   clusters out to redshifts of $\sim1.2$ in summed $r'+i'+z'$ SDSS
   data. Redshift estimates can be derived from the SDSS magnitudes of
   brightest cluster members out to redshifts near unity. Over the
   area of sky it covers, SDSS should find $\ga98\%$ of the clusters
   detectable by the {\em Planck\/} satellite through the thermal
   Sunyaev-Zel'dovich effect. The few {\em Planck\/} clusters not
   detected in SDSS will almost all be at $z\ga1.2$.}

\ifjournal
  \abstract{\abstext}
\else
  \begin{abstract}\abstext\end{abstract}
\fi

\maketitle

\section{Introduction}

Galaxy clusters are not only interesting for studying galaxy
evolution; they have also become an important tool for understanding
the growth of cosmic structure and the cosmological framework in which
it occurs. The construction of observed galaxy cluster samples out to
redshifts unity and beyond is thus an important goal of current
observational cosmology. Traditionally, galaxy clusters have been
identified at optical wavelengths as regions on the sky where the
number density of galaxies sufficiently exceeds its mean. This
definition requires of course that the cluster galaxies be
individually detectable, setting an upper limit to redshifts at which
galaxy clusters can be found.

Dalcanton (1996) proposed that clusters could be detected as regions
on the sky where the surface brightness exceeds the average sky
brightness. Her suggested procedure consists of removing galaxies from
carefully flat-fielded images, smoothing the residual image with a
kernel whose width should approximately match the angular extent of
galaxy clusters at intermediate redshifts, and searching for peaks in
the smoothed surface-brightness distribution which sufficiently exceed
the noise level of the smoothed sky background.

Gonzalez et al.~(2001) successfully applied this technique to the Las
Campanas Distant Cluster Survey (LCDCS) data taken with the Las
Campanas Great Circle Camera (Zaritsky, Schectman \& Bredthauer 1996)
and constructed a catalog of 1073 groups and clusters. Estimated
redshift limits of the catalog range from $\sim0.3$ for groups to
$\sim0.8$ for massive galaxy clusters.

Being intrinsically highly uniform by construction, drift-scan surveys
like the LCDCS provide an optimal data basis for the application of
this cluster-detection technique. In this paper, we estimate the
redshift limits expected for the largest ongoing drift-scan survey,
the Sloan Digital Sky Survey (SDSS; York et al.~2000). Compared to the
LCDCS, the SDSS uses a mirror area larger by a factor of $6.25$, an
exposure time shorter by a factor of $3.6$, and a system of broad-band
filters rather than a single, very broad filter. On the whole, the
flux limit of the SDSS is expected to be roughly a factor of two below
that of the LCDCS, which should allow the detection of massive
clusters out to redshifts beyond unity. We study this expectation in
detail in this paper. We note that cluster catalogues constructed from
the SDSS by more traditional techniques (i.e.~based on galaxy
catalogues) are already expected to be complete to $z\sim0.4$ (Kim et
al.~2001).

Sect.~\ref{sec:2} describes our assumptions. Results are presented in
Sect.~\ref{sec:3}, and we summarise our conclusions in
Sect.~\ref{sec:4}.

\section{Assumptions\label{sec:2}}

Throughout, we adopt a flat, low-density cosmological model with
matter density parameter $\Omega_0=0.3$, cosmological constant
corresponding to density parameter $\Omega_\Lambda=0.7$, and a Hubble
constant $H_0=70\,{\rm km\,s^{-1}\,Mpc^{-1}}$. The CDM power spectrum
is taken to have primordial spectral slope $n=1$ and is normalised
using $\sigma_8=0.93$ so as to reproduce the local abundance of galaxy
clusters.

\subsection{The SDSS photometric system}

\begin{figure}[ht]
  \includegraphics[width=\hsize]{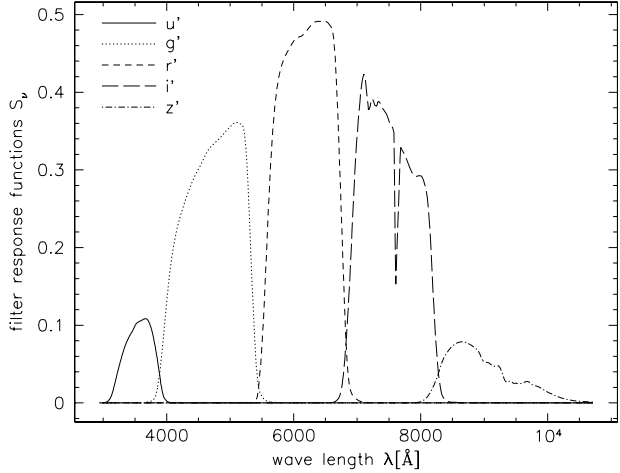}
\caption{Recently re-measured SDSS filter response functions. From
left to right: $u'$, $g'$, $r'$, $i'$, and $z'$.}
\label{fig:1}
\end{figure}

The SDSS photometric system uses five wide, almost non-overlapping
bands covering the wavelength range between 3000\,\AA\ and
11000\,\AA. The recently re-measured filter response functions $S_\nu$
are shown in Fig.~\ref{fig:1} (M.~Strauss, private communication). The
effective wavelengths $\lambda_\mathrm{eff}$, effective band widths
$\Delta\lambda_\mathrm{eff}$, and flux sensitivity quantities
\begin{equation}
  Q=\int\d(\ln\nu)\,S_\nu
\label{eq:1}
\end{equation}
of the five bands are summarised in Tab.~\ref{tab:1}.

\begin{table}[ht]
\begin{center}
\begin{tabular}{|l|rrrrr|}
\hline
band & $u'$ & $g'$ & $r'$ & $i'$ & $z'$ \\
\hline
$\lambda_\mathrm{eff}$ [\AA] &
  $3546$ & $4670$ & $6156$ & $7471$ & $8918$ \\
$\Delta\lambda_\mathrm{eff}$ [\AA] &
  $457$ & $928$ & $812$ & $893$ & $1183$ \\
$Q$ &
  $0.0171$ & $0.0893$ & $0.0886$ & $0.0591$ & $0.0099$ \\
$\mu_\mathrm{sky}$ &
  $21.8$ & $21.3$ & $20.5$ & $19.5$ & $18.3$ \\
counts &
  $470$ & $3930$ & $8140$ & $13630$ & $6897$ \\
$\Delta\mu_\mathrm{sky}$ &
  $23.4$ & $24.0$ & $23.5$ & $22.9$ & $21.4$ \\
\hline
\end{tabular}
\end{center}
\caption{Effective wavelengths $\lambda_\mathrm{eff}$, effective band
widths $\Delta\lambda_\mathrm{eff}$, and flux sensitivity quantities
$Q$ of the five SDSS bands. Also listed is the sky brightness in the
five SDSS photometric bands at the SDSS telescope site at Apache
Point, and corresponding counts. The last row gives the 5-$\sigma$
Poisson fluctuation level of the sky background. $\mu_\mathrm{sky}$
and $\Delta\mu_\mathrm{sky}$ are given in magnitudes per square arc
second.}
\label{tab:1}
\end{table}

The system is described in detail by Fukugita et al.~(1996). Its zero
points are placed on the spectrophotometric $AB$ magnitude system, so
that magnitudes can directly be converted to fluxes in physical
units. Given the spectral flux $F_\nu$ in units of ${\rm
erg\,cm^{-2}\,s^{-1}\,Hz^{-1}}$, the broad-band $AB$ magnitude is
defined as
\begin{equation}
  m=-2.5\log_{10}\frac{\int\d(\ln\nu)\,F_\nu\,S_\nu}{Q}-48.60\;,
\label{eq:2}
\end{equation}
with $Q$ defined in Eq.~(\ref{eq:1}).

In order to assess the noise level of the sky background, we need to
relate the incoming flux to the detector counts. The number of
photoelectrons released by the incoming spectral flux $F_\nu$ is given
by
\begin{equation}
  N_\mathrm{e}=A\,t\,h^{-1}\,\int\d(\ln\nu)\,F_\nu\,S_\nu\;,
\label{eq:3}
\end{equation}
where $h$ is Planck's constant. The survey telescope area is
$A=4.9\times10^{4}\,\mbox{cm}^2$, the effective exposure time per band
is $t=54.1\,\mbox{s}$.

The main sources of noise are the count fluctuations of the sky
brightness and the brightness fluctuations in galaxies just below the
SDSS detection limit. For the sky brightness $\mu_\mathrm{sky}$ at the
SDSS telescope site at Apache Point, we use values measured by
M.~Richmond (see Tab.~\ref{tab:1}, given in magnitudes per square arc
second). Given the area $A$ of the SDSS survey telescope, the
effective exposure time $t$, and the SDSS response functions, the
brightness of the night sky can be converted to counts, which are
listed for each band in Tab.~\ref{tab:1}. The Poisson fluctuation
level of the night sky per square arc second is on the order of a few
per cent. The 5-$\sigma$ Poisson fluctuation in magnitudes per square
arc second is given in the last row of Tab.~\ref{tab:1}.

The noise contributed by undetected galaxies is small compared to the
shot noise from the sky brightness. Because of the flatness of their
number-count function, most of the light from undetected galaxies
comes from galaxies just below the detection limit. Suppose the
typical solid angle covered by such a galaxy is $A$, and their number
density per solid angle is $n_\mathrm{gal}$. Let the photon count
produced per solid angle of the night sky be $n_\mathrm{e, sky}$, and
assume that the detection threshold is $\nu$ times the sky brightness
fluctuations. The number of sky photons per galaxy is then
$(n_\mathrm{e, sky}A)$, and the detection threshold is
$\nu(n_\mathrm{e, sky}A)^{1/2}$. The surface brightness from galaxies
at the detection threshold is then
\begin{equation}
  S_\mathrm{gal}=n_\mathrm{gal}\,\nu(n_\mathrm{e, sky}A)^{1/2}\;,
\label{eq:4}
\end{equation}
and its Poisson fluctuation in the solid angle $\delta\Omega$ is
\begin{equation}
  \Delta S_\mathrm{gal}=(n_\mathrm{gal}\delta\Omega)^{1/2}\,
  \nu(n_\mathrm{e, sky}A)^{1/2}\;.
\label{eq:5}
\end{equation}
The Poisson fluctuation in the sky brightness is
\begin{equation}
  \Delta S_\mathrm{sky}=(n_\mathrm{e, sky}\delta\Omega)^{1/2}\;.
\label{eq:6}
\end{equation}
This implies that the Poisson fluctuations in the flux from undetected
galaxies are small compared to the sky brightness fluctuations, since
the ratio
\begin{equation}
  \frac{\Delta S_\mathrm{gal}}{\Delta S_\mathrm{sky}}=
  \nu(n_\mathrm{gal}A)^{1/2}
\label{eq:7}
\end{equation}
scales with the square root of the sky fraction covered by galaxies,
which is on the per cent level.

The variance in the flux from undetected galaxies exceeds the Poisson
expectation because the galaxies are correlated. The contribution from
the correlation is, however, smaller by approximately a factor of ten
than the Poisson fluctuations at the apparent magnitude levels of
relevance (e.g.~Brainerd, Smail \& Mould 1995), so that the total
fluctuations expected from undetected galaxies are negligible compared
to the sky brightness fluctuations.

\subsection{Cluster spectra}

We describe the galaxy cluster spectrum $F_\nu^\mathrm{C}$ as a
weighted superposition of an early- and a late-type galaxy spectrum,
$F_\nu^\mathrm{E}$ and $F_\nu^\mathrm{L}$, respectively. The combined
cluster spectrum at redshift $z$ is then written as
\begin{eqnarray}
  F_\nu^\mathrm{C}(z)&=&f_E(z)\,l_E(z)\,F_{\nu(1+z)}^\mathrm{E}
  \nonumber\\
  &+&[1-f_E(z)]\,f_L\,F_{\nu(1+z)}^\mathrm{L}\;.
\label{eq:8}
\end{eqnarray}
The early-type fraction of the cluster population, $f_E(z)$, decreases
with redshift. The results of van Dokkum et al.~(2000) suggest
choosing a linear decrease with redshift,
\begin{equation}
  f_E(z)=0.8-0.4\,z\;.
\label{eq:9}
\end{equation}
The mass-to-light ratio of early-type galaxies in clusters decreases
with increasing redshift. For the mass-to-light ratio in the $B$ band,
van Dokkum et al.~(1998) find $\Delta\log(M/L_B)\sim-0.4\,z$. Adopting
this relation, we assume that early-type galaxies brighten with
redshift as
\begin{equation}
  l_E(z)=10^{0.4\,z}\;.
\label{eq:10}
\end{equation}
The late-type fraction increases with redshift, but we assume the
mass-to-light ratio of late-type galaxies to be constant. In order to
account for the fact that late-type galaxies are typically fainter
than early-type galaxies, we introduce a redshift-independent factor
$f_L<1$, which, based on the data of van Dokkum et al.~(2000), we set
to
\begin{equation}
  f_L=\frac{2}{3}\;.
\label{eq:11}
\end{equation}
The spectra $F_\nu^\mathrm{E,L}$ themselves are redshifted, but
otherwise assumed to be constant. They were kindly provided by
S.~Charlot (private communication).

The mass-to-light ratio of galaxy clusters is typically measured to be
on the order of $M/L_B=250\,M_\odot/L_\odot$ in the $B$ band
(e.g.~Bahcall, Lubin \& Dorman 1995; Carlberg et al.~1996; Carlberg,
Yee \& Ellingson 1997a; Gonzalez et al. 2000), however with a scatter
of $\sim20\%$. We normalise the combined cluster spectrum such that
the $B$-band mass-to-light ratio at redshift zero is $250$ in solar
units. Specifically, let $F_\nu^\odot$ be the solar spectrum, then the
amplitude of the cluster spectrum is chosen such that
\begin{equation}
  \int\d(\ln\nu)\,F_\nu^\mathrm{C}\,S_{\nu,B}=
  \frac{M}{250\,M_\odot}\,\int\d(\ln\nu)\,F_\nu^\odot\,S_{\nu,B}
\label{eq:12}
\end{equation}
is satisfied. The mass-to-ratios in all five SDSS bands and in the
conventional Johnson-$UBVRI$ bands are given in Tab.~\ref{tab:2}.

\begin{table}
\begin{center}
\begin{tabular}{|l|rrrrr|}
\hline
SDSS band & $u'$ & $g'$ & $r'$ & $i'$ & $z'$ \\
$(M/L)/(M_\odot/L_\odot)$ & 362 & 229 & 162 & 134 & 101 \\
Johnson band & $U$ & $B$ & $V$ & $R$ & $I$ \\
$(M/L)/(M_\odot/L_\odot)$ & 363 & 250 & 186 & 145 & 103 \\
\hline
\end{tabular}
\end{center}
\caption{Mass-to-light ratios in solar units of a galaxy cluster at
redshift zero in the five SDSS bands. The mass-to-light ratio is
normalised to $250$ in the Johnson $B$ band. For comparison, the
mass-to-light ratios in the conventional Johnson bands are also
shown. Due to the predominantly red galaxy population, the
mass-to-light ratio decreases towards longer-wavelength bands.}
\label{tab:2}
\end{table}

The predominantly early-type galaxy population causes the
mass-to-light ratio to decrease towards longer-wavelength bands. This
trend agrees well with observations; for instance, the $I$-band
mass-to-light ratio of Abell~1651 was measured to be $\sim160$
(Gonzalez et al.~2000).

\begin{figure}[ht]
  \includegraphics[width=\hsize]{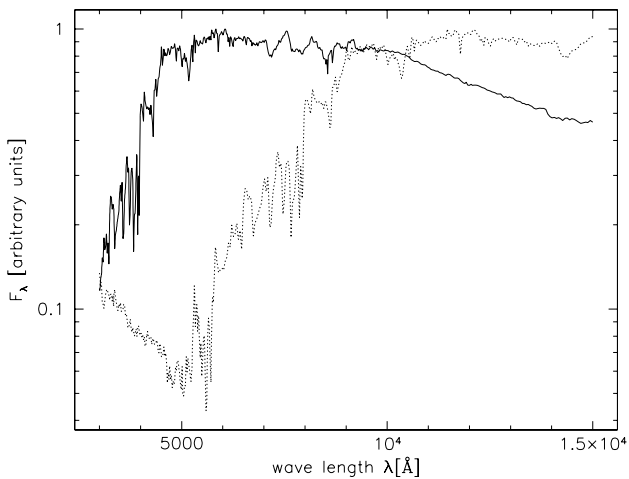}
\caption{Composite cluster spectra $F_\lambda^\mathrm{C}$ as functions
of wavelength $\lambda$ in \AA, at redshifts zero (solid curve) and
unity (dotted curve). Both spectra are arbitrarily normalised.}
\label{fig:2}
\end{figure}

Figure~\ref{fig:2} shows arbitrarily normalised cluster spectra at
redshifts zero and unity.

\subsection{Cluster $k$-corrections and colours}

\begin{figure}[ht]
  \includegraphics[width=\hsize]{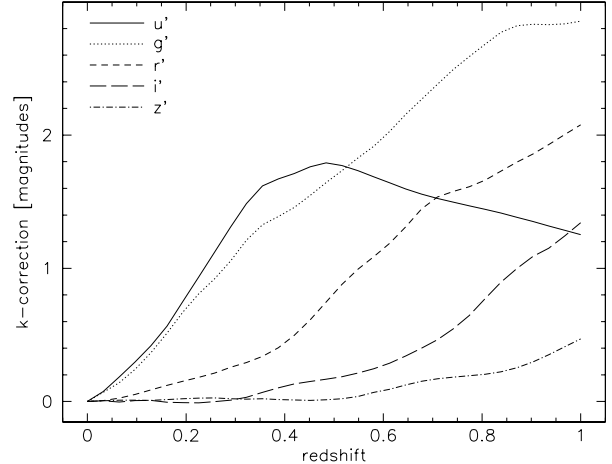}
\caption{Cluster $k$-corrections as a functions of redshift for
$M=5\times10^{14}\,M_\odot/h$ in the five SDSS bands (solid curve:
$u'$, dotted curve: $g'$, short-dashed curve: $r'$, long-dashed curve:
$i'$, dash-dotted curve: $z'$)}
\label{fig:3}
\end{figure}

Integrating the synthetic cluster spectrum with the SDSS filter curves
as in Eq.~(\ref{eq:2}), we can now compute $k$-corrections and cluster
colours for the SDSS photometric system. Figure~\ref{fig:3} displays
the cluster $k$-corrections in the five SDSS bands as functions of
cluster redshift.

The $k$-correction in the bluest SDSS band ($u'$) grows to
$\sim2.4$~mag.~up to $z\sim0.5$ and then levels off. This is because
at that redshift essentially all of the cluster flux has been shifted
redward of the $u'$ filter curve, and the remaining short-wavelength
part of the cluster spectrum is almost flat. The $k$ corrections in
the other SDSS bands grow monotonically with redshift. Their
amplitudes decrease as the filters get redder.

\begin{figure}[ht]
  \includegraphics[width=\hsize]{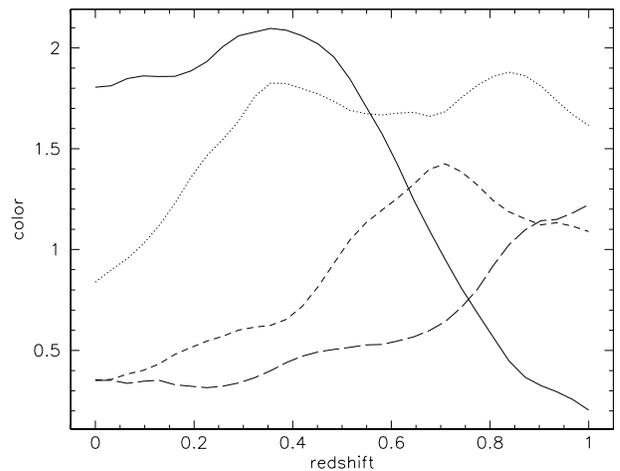}
\caption{Cluster colours as functions of redshift, for cluster mass
$M=5\times10^{14}\,M_\odot/h$. Solid curve: ($u'$-$g'$); dotted curve:
($g'$-$r'$), short-dashed curve: ($r'$-$i'$); long-dashed curve:
($i'$-$z'$).}
\label{fig:4}
\end{figure}

Figure~\ref{fig:4} shows the four different cluster colours
($u'$-$g'$), ($g'$-$r'$), ($r'$-$i'$) and ($i'$-$z'$) as a function of
cluster redshift.

On the whole, clusters become redder as their redshift increases, with
the exception of the bluest colour shown ($u'$-$g'$), which starts red
at low redshift and then decreases beyond redshift $z\sim0.4$ as the
3000\,\AA\ break shifts through the $g'$ band.

\subsection{Cluster surface brightness}

We assume that the average light distribution in galaxy clusters
follows mass, and that the mass distribution is described by the
familiar NFW density profile (Navarro, Frenk \& White 1996,
1997). Data from the CNOC survey shows that this gives a very good
model for the mean observed luminosity profile of rich clusters
(Carlberg et al.~1997b). The three-dimensional matter density is
\begin{equation}
  \rho(x)=\frac{\rho_\mathrm{s}}{x\,(1+x)^2}\;,
\label{eq:13}
\end{equation}
where $r_\mathrm{s}$ is a characteristic scale radius,
$\rho_\mathrm{s}$ is the density scale, and $x\equiv r/r_\mathrm{s}$.
Although two free parameters appear in (\ref{eq:13}), there is
effectively only one parameter to set because the scale radius and the
density scale are related. We choose the virial mass $M$ as the
remaining single free parameter and compute $r_\mathrm{s}$ and
$\rho_\mathrm{s}$ from it, following the prescription given in Navarro
et al.~(1997). The virial mass is
\begin{equation}
  M=4\pi\,\rho_\mathrm{s}\,r_\mathrm{s}^3\,
  \left[\ln(1+c)-\frac{c}{1+c}\right]\;,
\label{eq:14}
\end{equation}
where the concentration $c=r_{200}/r_\mathrm{s}$ is the ratio between
the virial radius and the scale radius.

Given the cluster mass, its luminosity $L$ is determined by the
mass-to-light ratio
\begin{equation}
  L=\frac{M}{(M/L)}\,L_\odot\;.
\label{eq:15}
\end{equation}
When projected along the line of sight, its surface mass density of an
NFW halo is $\Sigma(x)=\rho_\mathrm{s}\,r_\mathrm{s}\,f(x)$, with
\begin{equation}
  f(x)=\frac{2}{x^2-1}\,
  \left[1-\frac{2}{\sqrt{x^2-1}}\arctan\sqrt{\frac{x-1}{x+1}}\right]\;,
\label{eq:16}
\end{equation}
(see e.g.~Bartelmann 1996). Therefore, the surface brightness profile
of the cluster is
\begin{equation}
  S_\mathrm{C}(x)=\frac{L\,f(x)}{4\pi r_\mathrm{s}^2}\,
  \left[\ln(1+c)-\frac{c}{1+c}\right]^{-1}\;;
\label{eq:17}
\end{equation}
$S_\mathrm{C}(x)$ is the energy radiated by the cluster per unit
unit surface area per unit time.

Photon conservation, and the Etherington relation between
angular-diameter and luminosity distance then imply that the flux
received from a cluster at redshift $z$ per unit time, frequency,
detector area and solid angle is
\begin{eqnarray}
  S_\nu^\mathrm{C}(\theta)&=&\frac{F_\nu^\mathrm{C}(z)}
  {4\pi(1+z)^3\,4\pi r_\mathrm{s}^2}\,
  \left[\ln(1+c)-\frac{c}{1+c}\right]^{-1}\,
  \nonumber\\&\times&
  f\left[\frac{D(z)\theta}{r_\mathrm{s}}\right]\;,
\label{eq:18}
\end{eqnarray}
where $\theta$ is the angular separation from the projected cluster
centre, $D(z)$ is the angular diameter distance to the cluster, and
$F_\nu^\mathrm{C}$ is the cluster spectrum of Eq.~(\ref{eq:8}),
i.e.~the cluster luminosity per unit frequency.

By means of Eq.~(\ref{eq:2}), $S_\nu^\mathrm{C}(\theta)$ can now be
converted into a surface brightness profile for the cluster in the
conventional units of magnitudes per square arc second. The dotted
curve in Fig.~\ref{fig:6} shows an example for a cluster of mass
$2\times10^{14}\,M_\odot/h$ at redshift $z=0.75$.

\subsection{Smoothing}

\begin{figure}[ht]
  \includegraphics[width=\hsize]{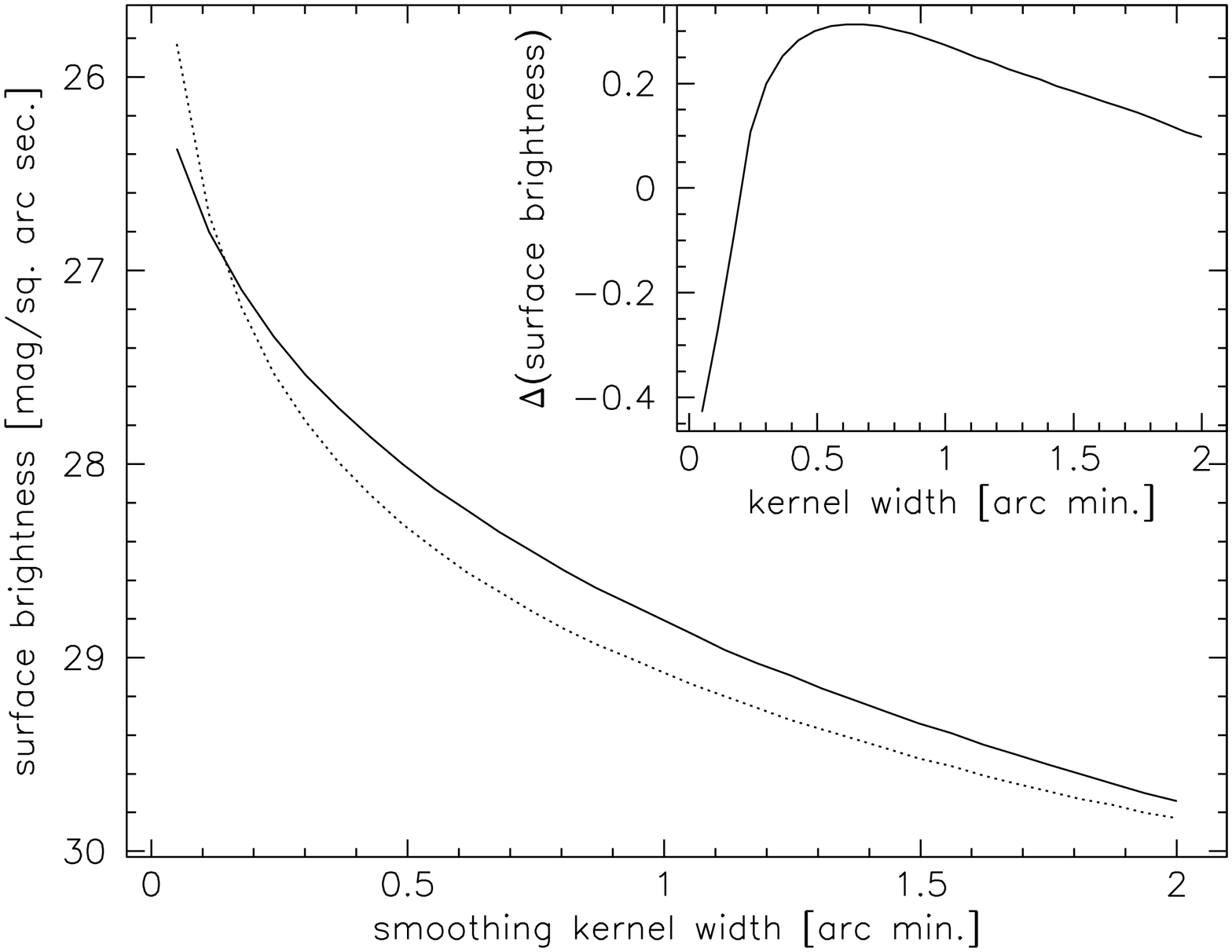}
\caption{Central surface brightness in the $r'$ band (in magnitudes
per square arc second) as a function of smoothing radius for a cluster
of mass $M=5\times10^{14}\,M_\odot/h$ at redshift $z=0.75$ (solid
curve) and the 5-$\sigma$ sky brightness fluctuations (dotted
curve). The small panel shows the difference between the curves.}
\label{fig:5}
\end{figure}

The signal-to-noise ratio of a cluster detection can be increased by
smoothing. We therefore convolve the projected cluster profile
(\ref{eq:18}) with a Gaussian filter of width
$\Delta\theta$. Exploiting the axial symmetry both of the cluster
profile and of the Gaussian, and making use of the convolution theorem
in Fourier space, the convolution can be transformed into
\begin{equation}
  \bar{S}_\nu^\mathrm{C}(\theta)=\int_0^\infty\d\phi\,\phi\,
  \frac{S_\nu^\mathrm{C}(\phi)}{\Delta\theta^2}\,
  \exp\left(-\frac{\theta^2+\phi^2}{2\Delta\theta^2}\right)\,
  \mathrm{I}_0\left(\frac{\theta\phi}{\Delta\theta^2}\right)\;,
\label{eq:19}
\end{equation}
where $\mathrm{I}_0(x)$ is the modified Bessel function of order
zero. It can be approximated by $\mathrm{I}_0(x)\approx(2\pi
x)^{-1/2}\,\exp(x)$ for $x\gg1$, in which case the convolution
simplifies to
\begin{equation}
  \bar{S}_\nu^\mathrm{C}(\theta)\approx\int_0^\infty\d\phi\,\phi\,
  \frac{S_\nu^\mathrm{C}(\phi)}{\sqrt{2\pi\theta\phi}\Delta\theta}\,
  \exp\left[-\frac{(\theta-\phi)^2}{2\Delta\theta^2}\right]\;.
\label{eq:20}
\end{equation}
We show in Fig.~\ref{fig:5} the central surface brightness in the $r'$
band of a cluster with mass $M=5\times10^{14}\,M_\odot/h$ at redshift
$0.75$, smoothed with a Gaussian kernel of varying width
$\Delta\theta$, and the 5-$\sigma$ sky brightness fluctuation
level. The small inserted panel shows the difference between the two
curves.

Figure~\ref{fig:5} shows that the sky noise falls below the central
cluster surface brightness when the smoothing kernel grows above
$\approx0.2'$, and the height of the cluster centre above the noise
reaches a broad maximum at $\Delta\theta\approx0.5'$. Correspondingly,
we choose a smoothing kernel width of $\Delta\theta=0.5'$ in the
following.

\begin{figure}[ht]
  \includegraphics[width=\hsize]{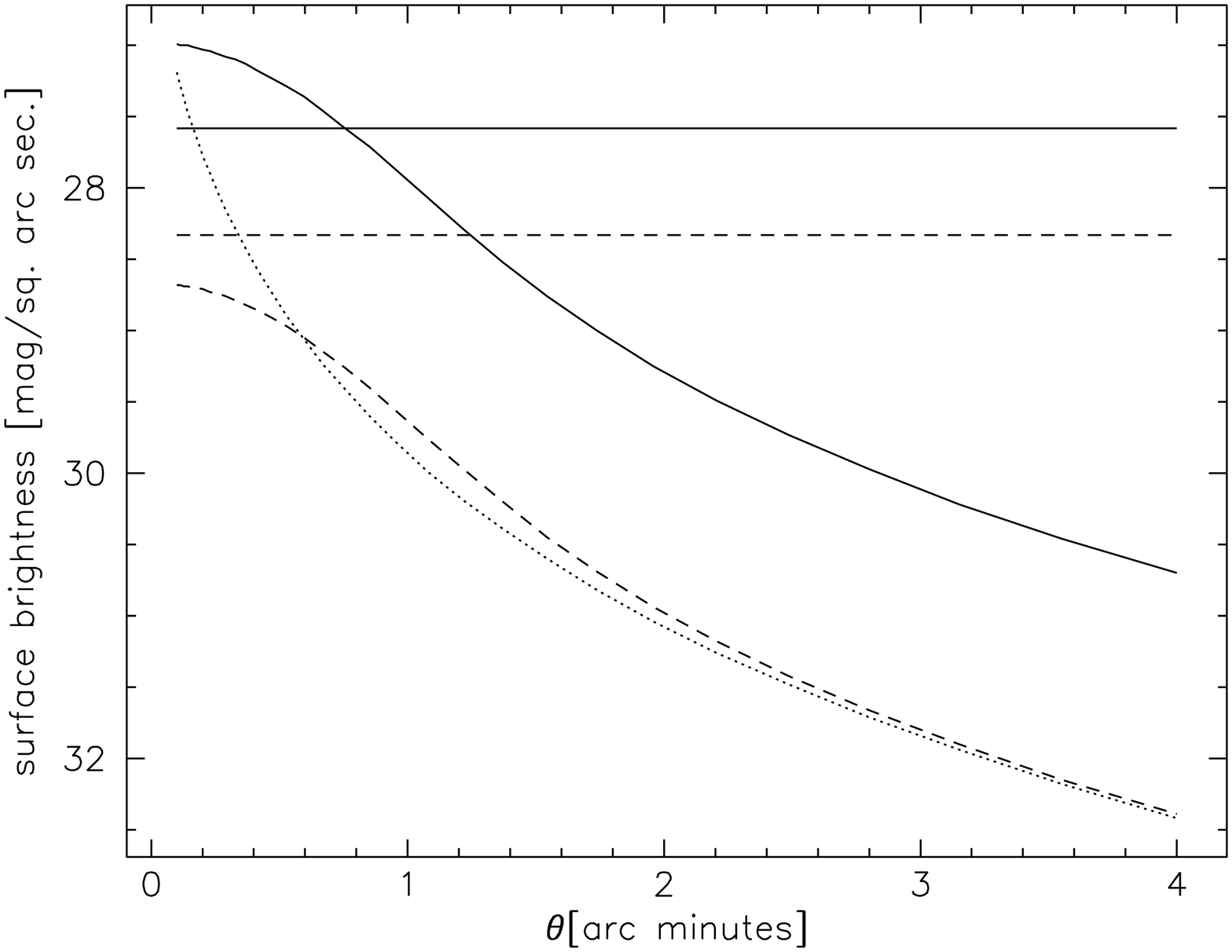}
\caption{Surface brightness profiles (in magnitudes per square arc
second) for a cluster of mass $M=2\times10^{14}\,M_\odot/h$ at
redshift $z=0.75$ as a function of cluster-centric distance
$\theta$. The assumed intrinsic NFW surface brightness profile in the
$r'$ band is shown as the dotted curve, which is smoothed by a
Gaussian kernel of width $\Delta\theta=0.5'$ (dashed curve). The solid
curve shows the surface brightness summed in the $g'$, $r'$ and $i'$
bands. The horizontal curves indicate the 5-$\sigma$ sky brightness
fluctuation levels in the $r'$ band (dashed) and the summed $g'$, $r'$
and $i'$ bands (solid), respectively. While the smoothed cluster falls
below the noise level in the $r'$ band, it is well detected in the
summed bands.}
\label{fig:6}
\end{figure}

Figure~\ref{fig:6} illustrates the effects of smoothing and sky noise
fluctuations on the surface-brightness profile of a cluster with mass
$M=2\times10^{14}\,M_\odot/h$ at redshift $z=0.75$. The intrinsic
profile is broadened by the Gaussian smoothing kernel, which has
$\Delta\theta=0.5'$. While the smoothed profile falls below the
5-$\sigma$ noise fluctuation level if data in the $r'$ band only are
used, the cluster is well detected if data in the $g'$, $r'$ and $i'$
bands are summed.

\section{Results\label{sec:3}}

\begin{figure}[ht]
  \includegraphics[width=\hsize]{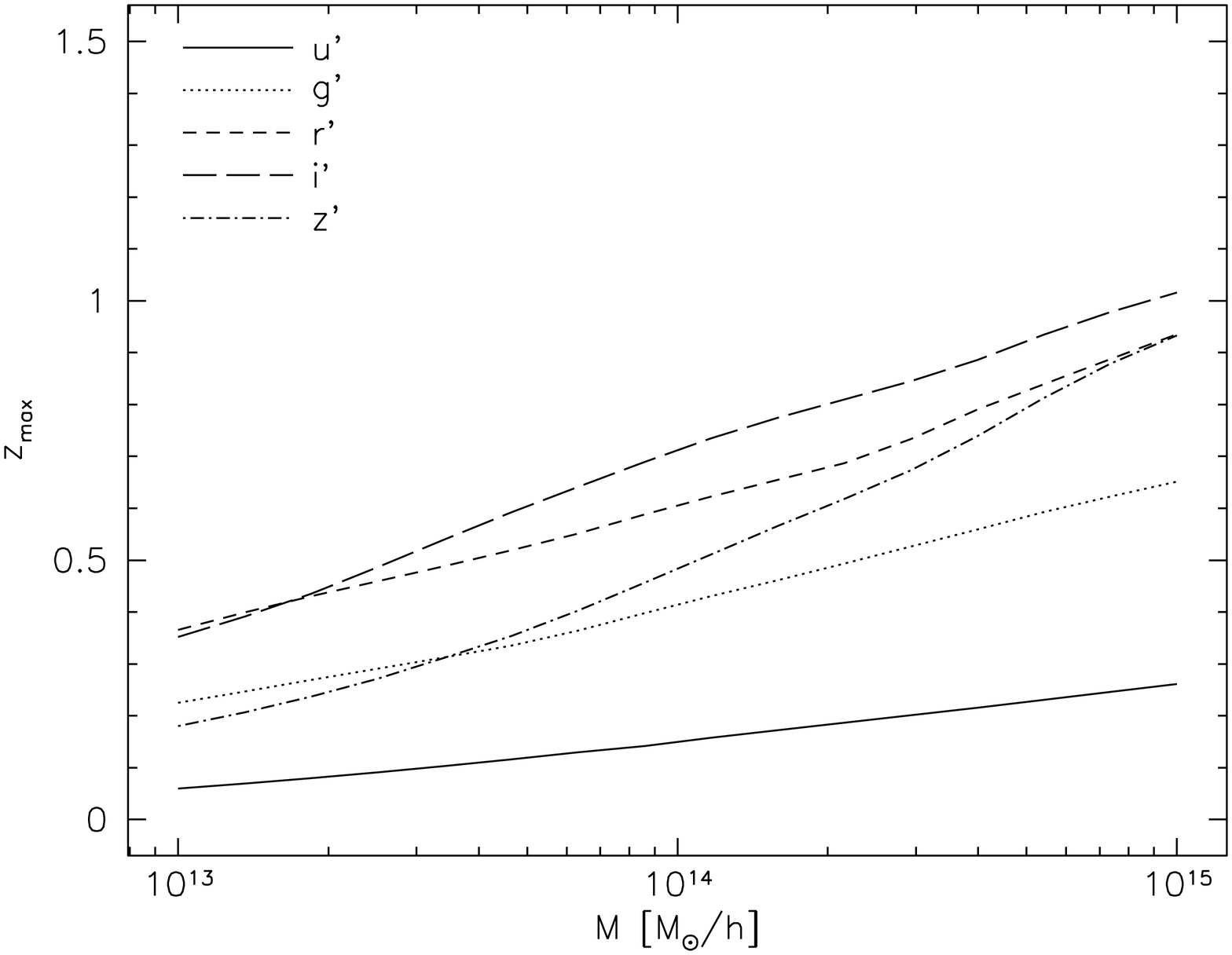}
\caption{5-$\sigma$ detection limits for clusters as a function of
mass in the five individual SDSS bands. Solid curve: $u'$, dotted
curve: $g'$, short-dashed curve: $r'$, long-dashed curve: $i'$,
dash-dotted curve: $z'$.}
\label{fig:7}
\end{figure}

We can now proceed to compute the upper redshift limit for a
significant cluster detection. We first consider individual SDSS
bands. Figure~\ref{fig:7} displays the redshift $z_\mathrm{max}$ as a
function of cluster mass at which the central cluster surface
brightness in each of the five SDSS bands drops below the 5-$\sigma$
limit of the sky background fluctuation, after smoothing with a
Gaussian kernel of width $\Delta\theta=0.5'$.

Clearly, the detection limit in the $u'$ band is the poorest, reaching
only out to redshift $z\sim0.15$ at $M=10^{14}\,M_\odot/h$. This is a
consequence of the combined effect of the relatively high background
fluctuation level, the low detector efficiency and the predominantly
red cluster light. The detection limit increases rapidly with
increasing filter wavelength to reach just above redshift unity in the
$i'$ band for massive clusters with $M=10^{15}\,M_\odot/h$. The
relatively low $k$ correction in the $z'$ band leads to the
comparatively steep increase in the upper redshift limit despite the
poor efficiency in the redmost SDSS band. 

\begin{figure}[ht]
  \includegraphics[width=\hsize]{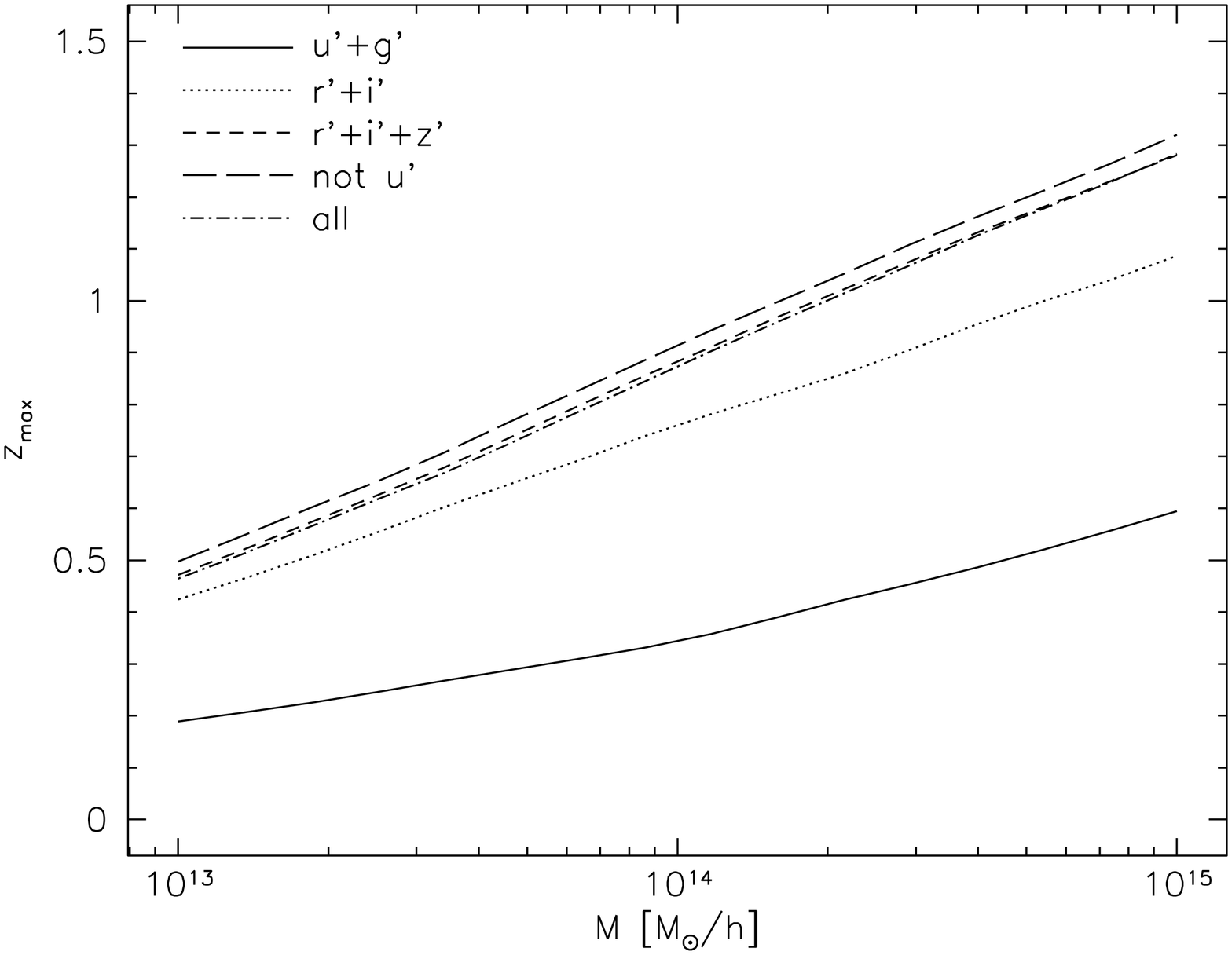}
\caption{5-$\sigma$ detection limits for clusters as a function of
mass in several combinations of wave bands. Solid curve: ($u'+g'$);
dotted curve: ($r'+i'$); short-dashed curve: ($r'+i'+z'$); long-dashed
curve: all bands except $u'$; dash-dotted curve: all bands.}
\label{fig:8}
\end{figure}

Figure~\ref{fig:7} indicates that the detection limit can be
considerably increased by summing data in several
bands. Figure~\ref{fig:8} quantifies this expectation.

The solid curve shows the detection limit for data summed in the two
bluest bands ($u'$ and $g'$), the dotted curve for data summed in the
two redder bands ($r'$ and $i'$). The increase between the two is
substantial; the upper redshift limit approximately doubles. One to
two tenths in redshift are gained if the $z'$ band data are added to
the $r'$ and $i'$ bands. Further addition of the $g'$ band improves
the limit a little more, while the relatively low signal-to-noise
ratio in the $u'$ band even lowers the detection limit if data in all
bands are summed.

These results show that massive clusters with
$M\sim10^{15}\,M_\odot/h$ should be detectable as significant
surface-brightness enhancements in the SDSS data out to redshifts of
$z\sim1.25$ if data in the three redmost bands are summed. The
redshift limit of the cluster detection increases approximately
linearly with $\log_{10}(M\,h/M_\odot)$.

This redshift limit agrees with expectations raised by the Las
Campanas Distant Cluster Survey (LCDCS; Zaritsky et al.~1997; Gonzalez
et al.~2001). The LCDCS used a 1-m telescope with an effective
exposure time of 194~s, while the SDSS has a 2.5-m telescope and an
effective exposure time of 54.1~s. The product of telescope area and
exposure time is therefore larger for the SDSS by 75\%. The LCDCS uses
the broad $W$ filter. It covers the wavelength range between 4600~\AA\
and 7300~\AA, which overlaps with the SDSS $r'$ filter and major
fractions of the $g'$ and $i'$ filters. The combined width of the
$g'+r'+i'+z'$ filters is approximately 4000~\AA, or $\sim50\%$ larger
than the width of the $W$ filter. Furthermore, adding $z'$ data helps
because of the predominantly red colour of galaxy clusters. Adopting
the same quantum efficiency for the SDSS and LCDCS CCDs, it seems
reasonable to assume that the combined $g'+r'+i'+z'$ SDSS data lower
the LCDCS flux limit by approximately a factor of
$(1.5\times1.75)^{1/2}\approx1.6$.

The LCDCS contains groups out to redshifts of $\sim0.3$, and massive
clusters out to redshifts of $\sim0.8$, while we estimate respective
redshift limits for SDSS to be $\sim0.5$ and $\sim1.3$. As
Eq.~(\ref{eq:18}) shows, the observed surface brightness
$S_\nu^\mathrm{C}$ scales with redshift roughly as $(1+z)^{-3}$. The
estimated improvement in the flux limit by $\sim1.6$ thus leads to an
expected increase in the redshift limit of
\begin{equation}
  z_\mathrm{max}^\mathrm{(SDSS)}\sim
  1.6^{1/3}(1+z_\mathrm{max}^\mathrm{(LCDCS)})-1
\label{eq:21}
\end{equation}
which agrees reasonably well with our direct estimates. We thus
confirm the speculation by Gonzalez et al.~(2001) that the SDSS
redshift limit for massive clusters could reach $\sim1.25$.

\begin{figure}[ht]
  \includegraphics[width=\hsize]{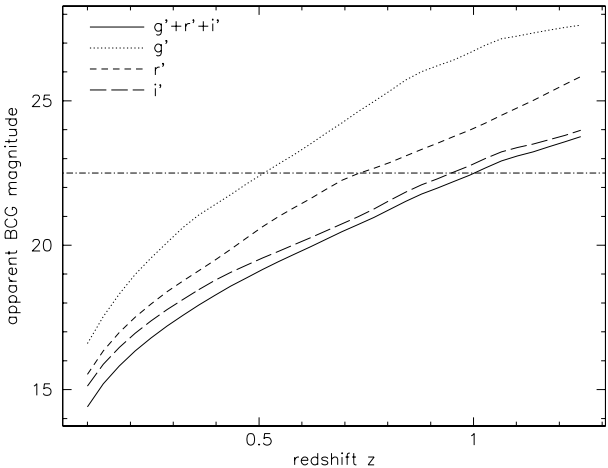}
\caption{Estimated magnitudes of brightest cluster galaxies in the
$g'$, $r'$ and $i'$ SDSS filter bands (dotted, short-dashed, and
long-dashed curves as indicated), and in the summed filters (solid
curve). The dash-dotted curve indicates a conservative detection limit
of $22.5$ magnitudes (approximately 5-$\sigma$). Brightest cluster
galaxies should be detectable in summed SDSS data out to redshifts
around unity. When identified with galaxy cluster candidates, they
will increase the reliability of the detection and allow photometric
redshift estimates.}
\label{fig:9}
\end{figure}

We note that it should be possible to estimate redshifts for a large
fraction of the SDSS cluster sample through the magnitudes of the
brightest cluster galaxies (BCGs). They form a remarkably homogeneous
class of objects (Hoessel, Gunn \& Thuan 1980; Schneider, Gunn \&
Hoessel 1983; Postman \& Lauer 1995) with a very narrow luminosity
function. The $K$-band absolute magnitudes of BCGs in the redshift
range $0\le z\le1$ are compatible with no luminosity evolution, and
their colour evolution is consistent with a passively evolving, old
stellar population (Arag\'on-Salamanca, Baugh \& Kauffmann
1998). Assuming a reduced absolute magnitude of $21.1$ in the $g'$
band (cf.~Schneider et al.~1983) and a non-evolving early-type
spectral energy distribution, we estimate the apparent BCG magnitudes
shown as functions of redshift in Fig.~\ref{fig:9}. We checked and
confirmed that they are compatible with the $K$-band measurements by
Arag\'on-Salamanca et al.~(1998)

If data in the $g'$, $r'$ and $i'$ data are summed, it should be
possible to identify BCGs out to redshifts near unity, assuming a
conservative detection limit of $22.5$ in the summed bands. Note that
the estimated $r'$ magnitudes out to redshifts of $\sim0.5$ agree very
well with the $r'$ magnitudes of galaxies with measured redshifts in
the SDSS Luminous Red Galaxy Sample (Eisenstein et al.~2001).

\begin{figure}[ht]
  \includegraphics[width=\hsize]{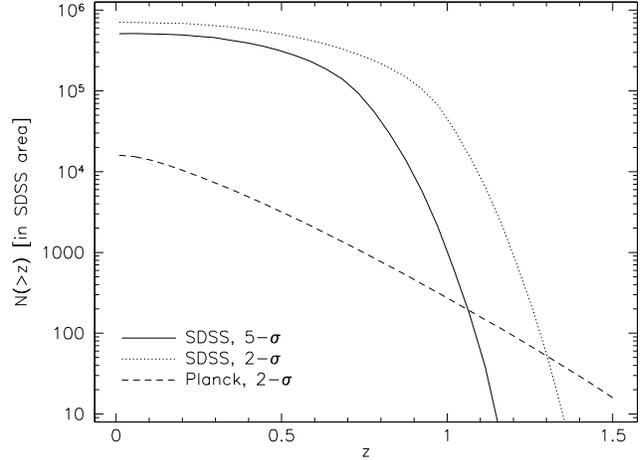}
\caption{Cumulative redshift distributions $N(>z)$ of clusters with
$M\ge5\times10^{13}\,h^{-1}\,M_\odot$ in summed $r'$+$i'$+$z'$ SDSS
data, detected at 2-$\sigma$ and 5-$\sigma$ significance as indicated
in the plot. Also shown is the cumulative redshift distribution of
clusters detected through the thermal Sunyaev-Zel'dovich effect at
2-$\sigma$ significance by the {\em Planck\/} satellite. Shown is the
expected number of clusters in the SDSS area. The abundance of SDSS
clusters lies almost two orders of magnitude above that for {\em
Planck\/} clusters up to the redshift where clusters with
$M=5\times10^{13}\,h^{-1}\,M_\odot$ drop below the detection
limit. The SDSS abundances then cut off exponentially. All {\em
Planck\/} clusters in the survey area are detected by SDSS out to
$z\sim1.1$ at 5-$\sigma$, and out to $z\sim1.3$ at 2-$\sigma$. Cluster
catalogues constructed from the SDSS data by more traditional
techniques are expected to be complete out to redshift $z\sim0.4$ (Kim
et al.~2001), and so should contain about 20\% of the clusters in
these deeper catalogues.}
\label{fig:10}
\end{figure}

Upcoming wide-area surveys in the sub-millimetre regime, like that
planned with the {\em Planck\/} satellite (Bersanelli et al.~1996),
will allow the detection of galaxy clusters through their peculiar
spectral signature caused by the thermal Sunyaev-Zel'dovich
effect. Taking the projected temperature sensitivity of {\em Planck\/}
and its characteristic beam width of $5'$ as a baseline and modelling
the cluster population with standard assumptions, Bartelmann (2001)
estimated the upper redshift limit of the cluster sample expected to
be detectable for {\em Planck\/}. Cumulative redshift distributions
for clusters with $M\ge5\times10^{13}\,h^{-1}\,M_\odot$ detected in
SDSS and {\em Planck\/} data are shown in Fig.~\ref{fig:10}.

As explained in Bartelmann (2001), {\em Planck\/} is expected to see
clusters at any redshift provided the cluster mass exceeds
$\sim5\times10^{14}\,h^{-1}\,M_\odot$. Below that mass threshold, the
cluster population expected to be detectable for {\em Planck\/} in the
SDSS survey area is completely contained in the SDSS cluster
sample. Since the cluster population is expected to die off rapidly at
redshifts beyond unity even in low-density cosmologies, it should be
possible to identify almost all ($\ga98\%$) {\em Planck\/} clusters in
the SDSS area with previously detected SDSS clusters. Thus SDSS will
provide positions, approximate redshifts and optical luminosities for
almost all clusters {\em Planck\/} will see over more than a third of
the usable high-latitude sky. {\em Planck\/} is expected to detect of
order $1.5$ galaxy clusters per square degree at the 2-$\sigma$
significance level. Therefore, it should be possible to identify
$\sim15000$ clusters in the SDSS data which are also detectable for
{\em Planck\/}, and approximately $70\%$ of those or $\sim11000$
clusters will be efficient weak lenses (Bartelmann 2001). The combined
data will allow detailed, multi-wavelength studies of a rich, uniquely
and homogeneously selected galaxy cluster sample. Correlation of the
two data sets (e.g.~requiring detection at $\ge2$-$\sigma$ in both or
stacking {\em Planck\/} data at the positions of SDSS clusters) will
allow one to go much deeper (cf.~Fig.~\ref{fig:10}). Note that the
{\em Planck\/} clusters which are {\em not\/} visible in SDSS should
almost all be at $z\ga1.2$, so that SDSS offers a way to identify the
small high redshift tail of the {\em Planck\/} cluster distribution.

\section{Summary and Conclusions\label{sec:4}}

Dalcanton (1996) suggested searching for galaxy clusters in optical
surveys by searching for excess surface brightness in heavily smoothed
images. A recent application of this technique, the Las Campanas
Distant Cluster Survey (Gonzalez et al.~2001) resulted in a catalog of
1073 groups and clusters with redshifts out to $\sim0.8$.

Drift-scan surveys like the LCDCS are ideal for this type of project
as they naturally yield highly uniform images. We investigated in this
paper redshift limits for cluster detection in the Sloan Digital Sky
Survey data. We assume that

\begin{itemize}

\item clusters have surface brightness profiles following the density
profile suggested by Navarro, Frenk \& White (1996, 1997);

\item the Universe is well described by a $\Lambda$CDM model
($\Omega_0=0.3$, $\Omega_\Lambda=0.7$, $h=0.7$) normalised to the
present number density of galaxy clusters;

\item the spectral energy distribution of clusters is a superposition
of a dominant early-type and a late-type galaxy spectrum;

\item the early-type fraction of galaxy clusters decreases with
increasing redshift, reaching half of its present value by redshift
unity;

\item early-type galaxies brighten with redshift; and

\item the mass-to-light ratio of galaxy clusters at present is
$250\,M_\odot/L_\odot$ in the Johnson $B$ band.

\end{itemize}

Then, using a measurement of the sky brightness at the SDSS telescope
site, applying the SDSS photometric system and smoothing with a
Gaussian kernel, we showed that:

\begin{itemize}

\item The most efficient single SDSS band for galaxy-cluster detection
is the $i'$ band, in which the 5-$\sigma$ detection limit for clusters
with mass $M\sim5\times10^{14}\,h^{-1}\,M_\odot$ is approximately
unity.

\item Summing data from different bands takes one considerably
further. While the $u'$ and $g'$ bands are not very efficient for this
purpose, summed $r'+i'+z'$ data allow clusters with
$M\sim5\times10^{14}\,h^{-1}\,M_\odot$ to be detected at 5-$\sigma$
significance out to redshift $\sim1.2$.

\item The limits derived here are in good agreement with
extrapolations from the Las Campanas Distant Cluster Survey.

\item Brightest cluster galaxies should be detectable in the SDSS data
out to redshifts near unity. For a substantial fraction of the SDSS
cluster sample, it will therefore be possible to derive redshift
estimates from the photometry of the brightest cluster members
associated with them.

\end{itemize}

Finally, it is worth noting that, except for the most massive
clusters, the SDSS cluster detection redshift limit falls above the
upper redshift limit for cluster detections in the sub-millimetre
regime expected with the upcoming {\em Planck\/} satellite. This
implies that an SDSS cluster sample should contain almost all the
clusters {\em Planck\/} is expected to see in the SDSS area, and that
it will thus become possible to study a sample of more than $10^4$
galaxy clusters in both the optical and sub-millimetre regimes, most
of which will be efficient weak gravitational lenses.

\newcommand{\acktext}
  {We wish to thank St\'ephane Charlot for providing us with spectral
   energy distributions, and Michael Strauss for updating the SDSS
   filter transmission curves.}

\ifjournal\acknowledgements{\acktext}\else
  \section*{Acknowledgements}\acktext
\fi

\end{document}